\documentstyle[12pt]{article}
\newcommand\ba{\begin{array}}
\newcommand\ea{\end{array}}
\newcommand\ben{\begin{equation}}
\newcommand\een{\end{equation}}
\newcommand\bea{\begin{eqnarray}}
\newcommand\eea{\end{eqnarray}}
\def\la{\mathrel{\mathpalette\fun <}}

\def\vev#1{{\langle #1 \rangle}}

\def\fun#1#2{\lower3.6pt\vbox{\baselineskip0pt\lineskip.9pt
\ialign{$\mathsurround=0pt#1\hfill##\hfil$\crcr#2\crcr\sim\crcr}}}

\def\math{\mathsurround 0pt}
\def\oversim#1#2{\lower.5pt\vbox{\baselineskip0pt \lineskip-.5pt

\ialign{$\math#1\hfil##\hfil$\crcr#2\crcr{\scriptstyle\sim}\crcr}}}

\def\lap{\mathrel{\mathpalette\oversim {\scriptstyle <}}}
\def\gap{\mathrel{\mathpalette\oversim {\scriptstyle >}}}

\def\({\left(} \def\){\right)}
\def\[{\left[} \def\]{\right]}
\def\pa{\partial}
\def\half{{\mathchoice{{\textstyle{1\over 2}}}{1\over 2}{1\over 2}{1
\over 2}}}
\def\unit#1{\ifinner \;
            \else \quad \fi
            {\rm #1}}

\def\phb{\bar\phi}
\def\vph{\varphi}
\def\vphi{\varphi}
\def\vphb{\bar\varphi}
\def\vphh{\hat\varphi}
\def\bx{\mbox{\boldmath $x$}}
\def\by{\mbox{\boldmath $y$}}
\def\bk{\mbox{\boldmath $k$}}
\def\tr{{\rm tr}}
\def\nab{\nabla}
\def\vphv{\varphi_v}
\def\vphbc{\bar\varphi_{\rm c}}
\def\seff{S_{\rm eff}}

\def\al{\alpha}
\def\be{\beta}

\def\de{\delta}

\def\la{\lambda}

\def\si{\sigma}

\def\om{\omega}
\def\Ga{\Gamma}
\def\De{\Delta}

\def\La{\Lambda}

\def\Om{\Omega}
\def\na{\nabla}

\begin{document}
\begin{titlepage}
\null\vspace{-62pt}
\begin{flushright}
DAMTP--93--28\\
Imperial/TP/92-93/40\\
{\tt hep-ph/9307253}\\
June 1993 \end{flushright}
\vspace{0.5in}
\baselineskip=36pt
\begin{center}
{{\LARGE \bf Thermal fluctuations at second order phase
transitions}}\\
\baselineskip=18pt
\vspace{1in}
{\large Mark Hindmarsh$^{(a,c)}$\\
\vspace{0.1in}
{\em and}\\
\vspace{0.1in}
R. J. Rivers$^{(b)}$\\}
\vspace{0.5in}
{{\it $^{(a)}$Department of Applied Mathematics and Theoretical
Physics,\\
University of Cambridge, Cambridge CB3 9EW}} \\
\vspace{.1in}
{{\it$^{(b)}$Blackett Laboratory, Imperial College, London SW7
2BZ}}\\
\end{center}

\vfill

\begin{abstract}
We construct a new coarse-grained effective potential which enables
us
to estimate the probabilities of thermal fluctuations above an
arbitrary threshold at different length scales.
\renewcommand{\thefootnote}{(\alph{footnote})}
\footnotetext[3]{ Address after October 1st 1993:
School of Mathematical and
Physical Sciences, University of Sussex, Brighton BN1 9QH, U.K.}
\end{abstract}
\end{titlepage}

\baselineskip=24pt

\section{Introduction}
\label{sIntro}
Quantum field theories with symmetry breaking display phase
transitions in
which, at high temperature, symmetry is restored.  Although the
critical temperatures and transition orders are calculable, using
the
techniques of hot field theory, the dynamics of transitions are
poorly
understood.

What is clear is that it is helpful to think in terms of the
fluctuations of
fields about their mean values.  Fluctuations can be classed as
`critical'
if they are extremal with respect to an appropriately defined energy
functional, `subcritical' if not.  Typical critical fluctuations are
the
bubbles invoked in bubble nucleation models of first order phase
transitions
\cite{langer} and the topological defects of second order
transitions \cite{cheetham}.
The role of subcritical fluctuations is less well understood, but
there
has been some interesting work by Gleiser, Kolb and others
\cite{g&k} suggesting that
they should not be ignored.  In this paper we attempt to lay the
groundwork for a fuller understanding of such fluctuations.

In the work of \cite{g&k} the quantities under study are the
probabilities
that fields will achieve fluctuations of specified types (e.g.
Gaussian
bubbles of false vacuum, or `emulsions' of such bubbles).  These are
sufficient to give us some indication as to whether the thermal
fluctuations
can cause the stable vacuum of a metastable system to be populated
prior
to quantum tunnelling.  However, precise calculations are needed for
reliable fluctuation rates and, while plausible, it is not obvious
that the
fluctuations that have been considered exhaust the relevant ones.
Furthermore, even if they are the relevant ones, the probabilities
in \cite{g&k} have been calculated by means of the {\it nonconvex}
one-loop effective potential.  For cold theories, a definition of
the effective potential in terms of probabilities
stresses its {\it convex} nature \cite{jona}, and we would expect a
similar situation
to apply here.

We adopt a different approach from \cite{g&k}, extending the work of
\cite{jona} to hot fields.
The probability
functional can be used directly to calculate the probability that a
single scalar
field exceeds an arbitrary threshold in some freely chosen region in
space.
We will find that a convex coarse-grained effective potential
emerges
naturally from this approach, which converges to the usual
definition
in the infinite volume limit.  A condensed matter
theorist would find this natural, but a field theorist may not.

We begin, as always, with the free field in Section \ref{sFree},
introducing the probability functional.  In Section \ref{sBounds} we
obtain our stated objective: an upper bound on the probability that
a
fluctuation exceed a threshold in some spatial volume.  We go on in
Section \ref{sIntSym} to calculate the upper bound for an
interacting
theory in its symmetric phase.  As might be anticipated, we are
presented with
difficulty in the broken phase, which we discuss in Section
\ref{broken}.

\section{Free field fluctuations}
\label{sFree}
To demonstrate the general methods, consider a
free field, with Euclidean classical action
\ben
S_4[\phi] = \int d^4 x\[\half(\pa \phi)^2 + \half m_0^2\phi^2\].
\een
At temperature $T=\be^{-1}$ (in units in which $k_{\rm B} =1$) the
partition
function can be written
\ben
Z = \tr \rho = \int D\vphb P[\vphb],
\een
where $P[\vphb] = \rho[\vphb,\vphb]$, the relative probability that
the
field
takes a specified configuration $\vphb(\bx)$. Expressed as a path
integral
the probability functional P is given by
\ben
P[\vphb] = \int_BD\phi \exp({-S_4[\phi]}),
\label{Pee}
\een
where the integral runs over field configurations periodic in
imaginary
time $\tau$ with period $\be$, subject to the boundary condition
\ben
B: \phi(\bx,\tau=0) = \vphb(\bx).
\een
The boundary condition is most simply enforced by introducing a
3-dimensional Lagrange multiplier field $\La(\bx)$, whereby
\ben
P[\phb] = \int D\La\int D\phi \exp\( -S_4[\phi] + i\be\int d\bx
\La(\bx)[\phi(\bx,0)-\vphb(\bx)]\).
\label{eProFun1}
\een
The periodicity in imaginary time is made manifest by the mode
expansion
\ben
\phi(\bx,\tau) = \sum_{n=-\infty}^{\infty} \vphi_n(\bx)e^{2\pi
in\tau/\be}.
\label{eModExp}
\een
At high $T$ the $n=0$ mode is the `light' mode, of mass $m_0$, and
the rest are
the `heavy' modes, with masses $m_n = \surd[m_0^2 + (2\pi n T)^2]$.
 If the
expansion (\ref{eModExp}) is inserted into (\ref{eProFun1}), the
integration
over heavy modes can be performed exactly to give
\ben
P[\vphb] = \int D\vphi_0 \exp\({-\be S_3[\vphi_0]}\)
\int D\La \exp\({i\be\int\La(\vphi_0-\vphb)-\half\be\int\La
K^{-1}\La}\).
\label{eProFun2}
\een
In (\ref{eProFun2}), $K^{-1}$ is the sum of the heavy mode
propagators
\ben
K^{-1} = \sum_{n\ne 0} (-\nab^2 + m_n^2)^{-1},
\label{eHeaPro}
\een
and $S_3[\vphi_0]$ is the spatial action
\ben
S_3[\vphi_0] = \int d\bx\[\half(\nab\vphi_0)^2 + \half
m_0^2\vphi_0^2\].
\een
On shifting the field to $\eta(\bx)=\vphi_0(\bx)-\vphb(\bx)$ in
(\ref{eProFun2}), the $\La$ integration can be performed to give
\ben
P[\vphb] = \int D\eta p[\eta]\exp\({-\be S_3[\vphb+\eta]}\),
\label{eProFun3}
\een
where $p[\eta]$ is the (normalised) thermal fluctuation distribution
functional
\ben
p[\eta] = N \exp\({-\half\be\int\eta K \eta}\).
\een
If follows that $Z$ can be written as
\bea
Z &=& \int D\vphb\int D\eta p[\eta] \exp\({-\be
S_3[\vphb+\eta]}\)\nonumber\\
  &=& \int D\vphb\exp\(-\be S_3[\vphb]\),
\eea
by virtue of the translational invariance of the formal measure.

We see that $\exp\(-\be S_3[\vphb]\)$ should not be identified with
the
relative probability distribution.  For the free field the
integration in
(\ref{eProFun3}) can be performed, to give
\ben
P[\vphb] = N\exp\(-\be H[\vphb]\),
\een
where $N$ is an undetermined normalisation constant, and the
`Hamiltonian'
$H$ is
\ben
H[\vphb] = S_3[\vphb] - \frac{1}{2}\int\({\de S_3\over \de\vphb}\)
(K+Q)^{-1}\({\de S_3\over \de\vphb}\).
\een
$K^{-1}$ has already been displayed in (\ref{eHeaPro}), and
\ben
Q^{-1} = (-\nab^2+m_0^2)^{-1}
\een
is the light mode propagator.  At high temperatures, $(K+Q)^{-1}
\simeq
K^{-1} \simeq  \be^2/12$, so that
\ben
H[\vphb] \simeq S_3[\vphb] - \frac{1}{24}\be^2\int\({\de S_3\over
\de\vphb}\)^2
 + O(\be^4).
\label{eHamExp}
\een
The second term in this equation should evince no surprise.  For the
simple
harmonic oscillator with potential $V(x) = \half m\om^2x^2$ the
exact
probability exponent is well known to be \cite{feyn}
\ben
H(x) = \frac{1}{2} m\om^2x^2\({2\tanh\be\om/2\over\be\om}\).
\label{eExaOsc}
\een
As calculated from (\ref{eHamExp}),
\bea
H(x) &=& V(x) - \frac{1}{24}\be^2\({dV\over dx}\)^2 + \cdots
\nonumber\\
     &=& \frac{1}{2} m\om^2x^2 - \frac{1}{24}\be^2(m\om^2x)^2 +
\cdots,
\eea
which is identical to the first two terms in the series expansion of
(\ref{eExaOsc}).

As an estimate of $H[\vphb]$ for free field fluctuations, consider
Gaussian fluctuations of the form \cite{g&k}
\ben
\vphb(\bx) = \vphi \exp(-\bx^2/2l^2)
\een
with constant $\vphi$.  Then, defining $\bar l = m_0 l$, we find
\ben
\beta H[\vphb] = \vphi^2/Am_0T
\een
where $A$ is the dimensionless quantity
\ben
A \simeq {2\over 3\pi^{3/2}}{1\over\bar l}\[1+{5\over
12}{1\over\bar{l}^2}
\({m_0\over T}\)^2 + \cdots \]
\label{ePhiSqu}
\een
The probability $P[\vphb]$ is exponentially damped for fluctuations
of magnitude $\vphi^2 \gg m_0T$ when $\bar l  = O(1)$.  In that case
the correction terms are $O(m_0^2/T^2)$, as expected.  Only for
smaller fluctuations of size $l=O(\be)$ is the second term
important.
The fact that $\vev{\vphb^2} = O(m_0T)$ has been discussed elsewhere
\cite{dine} and is
what we would expect if, in the field correlation functions, we cut
off the momenta at a scale $l^{-1}$.

\section{Gaussian probability bounds}
\label{sBounds}
Rather than pursue the likelihood of observing specific
fluctuations,
we return to $Z$, written as
\ben
Z = \int D\vphb \exp\(-\be H[\vphb]\).
\een
Since the integrand is a Boltzmann distribution for the field
probabilities
(unlike $e^{-\be S_3}$) this expression can be interpreted as the
partition function for a {\it classical\/} spin system with
Hamiltonian
$H[\vphb]$.  As such we can borrow the results of classical
statistical
mechanics.  In particular, we are interested in the following.
Consider a
volume $v$ in which the field takes an average value
\ben
\vphv = {1\over v} \int_{\bx\in v} d\bx \,\vphb(\bx).
\een
Let $p(\vphv\ge\vphi)$ be the probability that $\vphv\ge\vphi>0$.
Then
\ben
p(\vphv\ge\vphi) = {1\over Z} \int_{\vphv\ge\vphi} D\vphb
\exp\(-\be H[\vphb]\)
\een
It is straightforward to establish an upper bound for $p$ (see
\cite{jona}).  The Chebycheff
inequality gives ($j>0,\vphi>0$)
\bea
p(\vphv\ge\vphi)  &\le& {1\over Z} \int D\vphb \exp\(-\be H[\vphb] +
\be vj(\vphv-\vphi)\) \nonumber\\
                   &=& {1\over Z} \exp(-\be vj\vphi)
\int D\vphb \exp\(-\be H[\vphb] + \be\int d\bx J(\bx)\vphb(\bx)\)
\label{ch}
\eea
where $J(\bx) = jI(\bx)$, with $I$ the indicator (`window') function
\ben
I(\bx) = \left\{\ba{ll}
                    1 & \mbox{if $\bx\in v$} \\
                    0 & \mbox{if $\bx\not\in v$}.
                \ea\right.
\een
We now define a three dimensional generator of connected Greens
functions,
$W[J]$, by
\ben
W[J] = \be^{-1} \ln \[Z^{-1}
\int D\vphb\exp\(-\be H[\vphb] + \be\int d\bx J(\bx)\vphb(\bx)\)\],
\een
and an associated free energy $F$ which is its Legendre transform,
\ben
F[\vphbc] = -W[J] + \int J\vphbc \qquad\mbox{with}\qquad\vphbc(\bx)
=
{\delta W \over \delta J(\bx)}.
\een
Our bound then becomes minimised by
\bea
p(\vphv\ge\vphi) &\le& \left.\exp\(\be W[J] - \be\int
J\vphi\)\right|_{J=jI}
\nonumber\\
                 &=& \exp(-\be F[\vphi]).
\label{pbound}
\eea
with $j$ chosen such that $\int J(\bx)\vphbc(\bx)d\bx = \vphi jv$.

This can be evaluated exactly for a free field, for which $W[J]$
equals
\ben
W_0[J] = \frac{1}{2} \int d\bx d\by J(\bx)(K^{-1} +
Q^{-1})_{\bx\by}J(\by).
\label{Wzero}
\een
The two point Green
 function in this expression is just $\be\langle
\vphb(\bx)\vphb(\by)\rangle$, evaluated at $J=0$.  We wish to solve
\ben
\int J\left.{\delta W_0 \over \delta J}\right|_{jI} = \vphi jv
\een
for $j$.  This is easily done to obtain the Gaussian probability
bound
\ben
p(\vphv\ge\vphi) \le \exp\(-\half\vphi^2/\langle(\vphv)^2\rangle\).
\een
This bound is intuitively very appealing.  It says that the
probability
of fluctuations above a threshold $\vphi$ over a volume $v$ is
controlled
by the two point Green
 function averaged over that volume.  If we are
interested in coarse-grained fluctuations, we need only calculate
coarse-grained Green
 functions, and not worry about  high (spatial)
frequencies.  This result is actually well known in the theory of
Gaussian random fluctuations \cite{bardeen}, where it is often
expressed as the
fraction of space in which the fluctuations exceed the threshold
$\varphi$.

The evaluation of the coarse-grained Green
 function is very straightforward.
We find
\ben
\be v\langle(\vphv)^2\rangle  = {1\over v } \int {d\bk\over
(2\pi)^3}
|I(\bk)|^2 \sum_{n=-\infty}^{\infty}{1\over \bk^2 + m_n^2}
\een
where $I(\bk)$ is the Fourier transform of the indicator function.
Performing the sum in the usual way gives
\ben
\be v\langle(\vphv)^2\rangle  = {1\over v } \int {d\bk\over
(2\pi)^3}
|I(\bk)|^2{\be\over E} \[\frac{1}{2} +{1\over e^{\be E} - 1}\]
\label{hotint}
\een
where $E^2 = \bk^2 + m_0^2$.  In the high temperature limit ($\be
m_0
\ll 1$) we consider two regimes: $m^3v \gg 1$ and $m^3v \ll 1$.
For
large volumes, when $I(\bk)\rightarrow\delta(\bk)$, we find
\ben
\be v\langle(\vphv)^2\rangle \simeq {1\over m_0^2} (1+
O(\be^2m_0^2))
\label{greatv}
\een
That is,
\begin{equation}
p(\vphv\ge\vphi) \le \exp(-\half\be vm_0^2\vphi^2).
\een
while for small volumes (which are still large compared with
$\be^3$) we
essentially reproduce (\ref{ePhiSqu}), with $v \sim l^3$:
\ben
\be v\langle(\vphv)^2\rangle \sim
v^{2/3}(1-O(m_0v^{1/3},\be^2v^{-2/3})).
\label{miniv}
\een
The precise numerical factor here depends on the geometry of the
region.
As $v\to 0$ it is not hard to check that we recover the well known
result
$\langle\vphi^2\rangle = T^2/12$.

\section{Interacting fields:  Symmetric phase}
\label{sIntSym}
Now consider the scalar field $\phi$ with Euclidean action
\ben
S_4[\phi] = \int d^4x\[\frac{1}{2}(\pa\phi)^2
+\frac{1}{2}\mu_0^2\phi^2+\frac{1}{4!}\la
\phi^4\]
\een
The case of interest is $\mu_0^2 <0$, where the low temperature
ground state
breaks the $\phi \to -\phi$ symmetry.  As the temperature is raised,
the
symmetry is restored in a second order phase transition \cite{dj}.

We start by computing the probability functional $P[\vphb]$.
Inserting the new action into (\ref{Pee}), we find after integrating
out the
heavy modes in the Gaussian approximation that
\ben
P[\vphb] = \int D\phi_0 \exp\(-\be S_{\rm eff}[\phi_0]\)\int D\La
\exp\(i\be \int\Lambda(\vph_0 - \vphb) - \half\int \La K^{-1}\La\)
\label{eIntPro}
\een
where the two point Green
 function is now given by
\ben
K^{-1}[\vph_0] = \sum_{n\ne 0} \(-\nab^2 + m_n^2 +
\half\la\vph_0^2\)^{-1}.
\een
At high temperatures it is sufficient to take $K^{-1} \simeq
\be^2/12$, as
in (\ref{eExaOsc}).  The action in (\ref{eIntPro}) is the one-loop
effective action
\ben
S_{\rm eff} = \int d\bx\[\frac{1}{2}(\nab\vph_0)^2 + \frac{1}{2}
\mu_0^2\vph_0^2 +
\frac{1}{4!}\la\vph_0^4\] + \be^{-1}\sum_{n\ne 0}
\tr\ln(-\nab^2+m_n^2+\half\la\vph_0^2)
\label{fact}
\een
in which radiative corrections from the {\it heavy modes only\/}
have
been
taken into account.  We recall that
in an expansion in powers of $\mu_0/T$, the leading behaviour of
$S_{\rm eff}[\vph_0]$ is
\ben
S_{\rm eff}[\vph_0] = \int d\bx\[\frac{1}{2}(\nab\vph_0)^2 +
\frac{1}{2}\mu^2(T)\vph_0^2 +
\frac{1}{4!}\la\vph_0^4\]
\een
where
\ben
\mu^2(T)  = \mu_0^2 + \frac{1}{12}\la T^2.
\label{eMas}
\een
Since the first non-leading behaviour is only present in the
(unintegrated)
$n=0$ mode,  (\ref{eMas}) is correct to second non-leading behaviour
\cite{cheetham}.   With
$\mu_0^2 < 0$ there is a second order phase transition at a critical
temperature $T_{\rm c}$ where, in this approximation
\ben
T_{\rm c}^2 = -12\mu_0^2/\la.
\een

Let us first restrict ourselves to temperatures $T>T_c$, for
which $\mu^{2}(T)>0$.
We would perhaps like to have an expression for $P[\vph]$ of the
form
\ben
P[\vphb] = \int D\eta\, p[\eta]\exp\(-\be S_{\rm eff}[\vphb+\eta]\)
\een
just as for the free field (\ref{eProFun3}).
A little care is required however, for the interactions
affect
both the fluctuation probability functional $p$ and the effective
action $\seff$.

If we substitute our interacting action into the expression
(\ref{eProFun1}),
 then after performing  the heavy mode functional integrals we are
left
with
\ben
P[\vphb] = \int D\vph_0 D\La \exp\(-\be \seff[\vph_0] +
\be\Om[i\La,\vph_0]
+i\be\int\La(\vph_0-\vphb)\)
\label{eIntPro1}
\een
where $\Om[\{J_n\},\vph_0]$ generates connected Greens functions for
the heavy modes in a light mode background $\vph_0$:
\ben
\Om[\{J_n\},\vph_0] = -\be^{-1} \ln\[\int \prod_{n\ne 0}D\vph_{n}
\exp\(-\be S_3^n[\vph_n,\vph_0] + \be\int J_n\vph_n\)\].
\een
In (\ref{eIntPro1})  the generator $\Om$ is evaluated at current
$J_n = i\La$
for all $n$.  Extremizing with respect to $\La$, and shifting the
fields,
we define a probability distribution functional $\Gamma$,
so that
\ben
P[\vphb] = \int D\eta
\exp\(-\be\Gamma[\eta,\vphb+\eta]-\be\seff[\vphb+\eta]\)
\label{eIntPro2}
\een
where
\ben
\Gamma[\eta,\vph_0] = \Omega[i\La,\vph_0] - i\int\La\eta
\een
evaluated at $\de\Om/\de(i\La) = \eta$.   In the Appendix we show
that
$\Gamma$ has a high temperature perturbative expansion of the form
\ben
\Gamma[\eta,\vph_0] \simeq \frac{1}{2}\int\eta K(\vph_0)\eta +
\frac{1}{3!}\la
\int \eta^3 \vph_0 \si_a + \frac{1}{4!}\la\int\eta^4
\si_b
\een
where $\si_a$ and $\si_b$  are numerical constants.  Thus we find
that
the exponent in  (\ref{eIntPro2})  can be written
\ben
\Ga + \seff= \seff[\vphb] + \int{\de\seff\over\de\vphb}\eta +
\frac{1}{2}\int\eta[Q_{\rm eff}(\vphb)+K(\vphb)]\eta + {\rm
O}(\la\smallint\eta^3\vphb)
\een
where $Q_{\rm eff}$ is the one loop propagator of $\seff$.
A saddle-point evaluation of the functional integral over the
fluctuations
$\eta$ enables us to recover a `Hamiltonian'
\ben
H[\vphb] \simeq S_{\rm eff}[\vphb] - \frac{1}{24}\be^2 \int d\bx
\({\de S_{\rm eff}\over \de \vphb(\bx)}\)^2 + O(\be^4).
\label{eIntHam}
\een
Neglecting the interaction terms in the fluctuation
probability functional $\Ga$ is a good approximation, because
the corrections are of order $\la\be^6$.

Thus we are still entitled to write
\ben
P[\vphb] = N\exp\(-\be H[\vphb]\)
\label{pp}
\een
with $H$ the sum of a quadratic part $H_0$ and an interacting
part $H_{\rm I}$. As before, we wish to estimate the probability
$p(\vphv \ge \vph)$ that the field average in a volume $v$ exceeds
an
arbitrary threshold $\vph >0$. By the symmetry of the theory this is
equal to $p(\vphv \le -\vph)$. This gives us the possibility of a
perturbation expansion, since \ben \exp\(\be W[J]\) = \exp\(-\be
H_{\rm I}[\de/\de J]\)\exp\(\be W_0[J]\) \een where $W_0$ has
already been calculated in (\ref{Wzero}).   The $n$-point connected
Green
functions $G^{(n)}(\bx_1,\ldots,\bx_n)$ appear in the Taylor
expansion of $W$:
\ben
\be W[J] = \sum_n \frac{\be^n}{n!}\int d\bx_1\ldots
d\bx_nG^{(n)}(\bx_1,\ldots,\bx_n)J(\bx_1)\ldots J(\bx_n).
\een
Recall
that in order to minimize the upper bound on the probability we have
to solve \ben jv\vph = \int J\left.{\de W\over \de J}\right|_{jI}
\een for $jv$.  Thus we obtain a polynomial equation
\ben \sum_n
{1\over (n-1)!} (\be vj)^nG_v^{(n)} = \be vj\vph \label{ePoly}
\een where the
smeared Green
 functions $G_v^{(n)}$ are defined by \ben G_v^{(n)}
= \int d\bx_1\ldots d\bx_nG^{(n)}(\bx_1,.\ldots,\bx_n)I(\bx_1)
\ldots
I(\bx_n) / v^n = \langle(\vphv)^n\rangle \een The upper bound on the
probability is then obtained by exponentiating
\ben \sum_{n\ge
1}(\be vj)^nG_n^{(n)}/n! - \be vj\vph \een evaluated at the solution
to (\ref{ePoly}).

This we now proceed to do for our $\vph^4$ theory, to first order
in
the coupling $\la$.  To this order, we need only include terms up to
and
including the 4-point function, so that we have to solve
\ben
\frac{1}{3!}(\be vj)^3G_v^{(4)} + (\be vj)G_v^{(2)} = \vph.
\een
As usual, this can be done perturbatively around the zeroth order
solution
$(\be jv) = \vph/G_v^{(2)}$.  We find that the zeroth order solution
is shifted by
an amount $\vph^3G_v^{(4)}/(G_v^{(2)})^33!$, leading finally to
\ben
p(\vphv \le \vph) \le \exp\[-\frac{1}{2}\vph^2/G_v^{(2)} -
\frac{1}{4!}\vph^4
G_v^{(4)}/(G_v^{(2)})^4\].
\label{symbound}
\een
We have now evaluated the first two terms of a coarse-grained
effective
potential, which we call $V_v(\vph)$.  The general procedure for
calculating general terms can be guessed from the form of equation
(\ref{symbound}):
it is exactly like the generator of one-particle-irreducible
diagrams, but evaluated with smeared Green
functions.

It is convenient to write the bound (\ref{symbound}) as
\begin{equation}
p(\vphv\ge\vphi) \le \exp(-\be vV_v(\vphi)).
\label{vbound}
\een
in which the volume $v$ has been extracted from the exponent.
For
large volumes, when $I(\bk)\rightarrow\delta(\bk)$, we have
\ben
\be vG_{v}^{(2)}\simeq {1\over \mu^2 (T)}
\een
as in (\ref{greatv}).  Further,
\ben
G_v^{(4)}/(G_v^{(2)})^4\simeq \be\lambda v
\label{greg}
\een
whence
\ben
V_v(\vph)\rightarrow V(\vph) = \frac{1}{2}\mu^2 (T)\vph^2 +
\frac{1}{4!}\la\vph^4\
\label{vinf}
\een
the (one-loop) effective potential of the theory.

On the other hand, for small $v$, (but large enough to keep
$\be v^{-1/3}\ll 1$), we have
\ben
\be v G_{v}^{(2)}\sim v^{2/3}
\label{eSmallV}
\een
as in (\ref{miniv}).  However, (\ref{greg}) is still valid, as an
order of magnitude
result.
Thus $V_v(\vph)$ can be written as
\ben
V_v(\vph) = \frac{a}{2}\mu^2 (T)\vph^2 +
\frac{b}{4!}\la\vph^4\
\een
where
\ben
a^{-1}=O(\mu^2 (T) v^{2/3}),\qquad    b=O(1).
\een
We see that, qualitatively at least, $V_v(\vph)$ can be approximated
by its infinite volume limit $V(\vph)$ of (\ref{vinf}) for volumes
$v$
as small as correlation-size volumes $v=O(\mu^{-3}(T))$.  Only when
$v$ is significantly smaller is there a noticeable difference.
Thus on scales larger than the correlation length
we can think of the effective potential as
determining the probability bound.  As the phase transition is
approached, the correlation length diverges and (\ref{eSmallV})
takes over.

It is easy to see to what extent the bound is not Gaussian.  The
first term alone in $V(\vph)$ of (\ref{vinf}) would imply
\ben
\langle\vph_v^{2}\rangle=O(T\mu(T))
\label{otm}
\een
for correlation-sized volumes.
The $\la\vph^{4}$ exponent becomes important when
$\la\gap\be \mu(T)$ or, equivalently, when $T$ approaches
$T_c$.  This inequality,  written in the form
\ben
\(1-\frac{T^{2}}{T_c^{2}}\)\lap O(\la),
\label{Gin}
\een
was first discussed by Ginzburg \cite{Gin60}.
For temperatures lying so
close to $T_c$, the Gaussian approximation that led to (\ref{fact})
begins to break down.
We shall reencounter the Ginzburg inequality in a different guise
when we discuss the more interesting case of broken symmetry.

\section{Interacting fields: Broken symmetry phase}
\label{broken}
When $T<T_c$ the reflection symmetry is broken.  We are unable to
proceed as in the previous section since perturbation theory
is inappropriate for the construction of the broken symmetry
effective potential.  Nonperturbative analytic
approximations are not easy to find.  However, the relative
probabilities $P[\vphb]$ are still given by a Hamiltonian $H$, as in
(\ref{pp}). Since $H$ is positive definite, $P$ permits the bound
\ben
P[\vphb] = N\exp\(-\be H[\vphb]\) \leq N\exp\(-\be H_{v}[\vphb]\)
\label{ppp}
\een
where $H_{v}$ denotes the coarse-grained Hamiltonian obtained by
integrating the Hamiltonian density only over the volume $v$.
Our earlier
discussion has shown that the large-$v$ results are valid down to
correlation-volume size.

We can now proceed as before.  The Chebycheff inequality (\ref{ch})
is
still valid, but with all integrals restricted to $v$.  As in
(\ref{pbound}),
the probability bound is
\bea
p(\vphv\ge\vphi) &\le& \exp\(\be W_v[j] - \be\int_v j\vphi\)\\
                 &=& \exp(-\be F_v[\vphi]).
\eea
where
\ben
\be W_v[j] = \ln \[Z^{-1}
\int D\vphb\exp\(-\be H_v[\vphb] + \be j\int_v d\bx \vphb\)\].
\een
The second inequality in (\ref{ppp}) should not be significantly
weaker than
the first since our earlier
discussion has suggested that the large-$v$ results are valid down
to
correlation-volume size.

It is convenient to extract a volume $v$ in the expressions above.
That is, we write $F_v [\vphi]=v\bar{V}_v(\vphi)$ and
$W_v[j]=v\bar{\om}_v(j)$.
On integrating, $\bar{V}_v(\vphi)$ is the
Legendre transform of
\ben
\bar{\om}_v(j)=\om_v(j) + \frac{1}{24}\be^2j^2
\een
where $\om_v(j)$ is more simply defined by
\ben
\be v\om_v(j) = \ln \[Z^{-1}
\int D\vphb\exp\(-\be S_v[\vphb] + \be j\int_v d\bx \vphb\)\],
\label{bvo}
\een
in which
\ben
S_v[\vph_0] = \int_v d\bx\[\frac{1}{2}(\nab\vph_0)^2 + \frac{1}{2}
\mu^2(T)\vph_0^2 +
\frac{1}{4!}\la\vph_0^4\]
\een
the effective action restricted to $v$.

To first nonleading order in powers of $\mu/T$ we can neglect $\be^2
j^2/24$ provided we further restrict ourselves to volumes of linear
dimension $l\gg \be$, and this we shall assume.  By definition, with
$j$ chosen so that
\ben
\frac{d\om_v}{d j} = \vphi
\een
then
\ben
V_v(\vphi)=-\om_v(j)+j\vphi
\een
equals the effective potential in the infinite volume limit.  As a
Legendre transform it is, by definition, convex.

However, we can say more.  We adopt the tactics of \cite{or} (for
cold fields) and \cite{rr} (for hot fields) in writing
(\ref{bvo}) as
\ben
\exp\(\be v\om_v(j)\) = \int d\vph\int
D\vphb\de(\vph-\vphv)\exp\(-
\be S_v[\vphb]+\be v j \vph\).
\een
Further decomposing $\vphb(\bx)$ as $\vphb(\bx)= \vph+\eta(\bx)$
we find\footnote{Ref \cite{rr} was published without being
proofread, and may seem unclear at this point.}
\ben
\exp\(\be v\om_v(j)\) = \int d\vph\exp(\be vj\vph)\int
D\eta\de(\smallint_vd\bx\eta) \exp\(-\be S_v[\vph+\eta]\).
\label{eFunInt1}
\een
Let us recapitulate the results of the previous section for the
symmetric
phase, for which
\ben
V''(\vph) = \mu^2(T)+\frac{1}{2}\la\vph^2 > 0.
\een
A Gaussian approximation to (\ref{eFunInt1}) gives
\bea
&&\exp\(\be v\om_v(j)\) =\nonumber\\
&& \int d\vph\exp\(\be v[j\vph-V(\vph)]\)
\int D\eta\de(\smallint_vd\bx\eta) \exp\(-\be\int_v\eta\,\de
S_v/\de\vph-\half\be
\int_v\eta\bar{K}\eta\),
\label{eGauApp}
\eea
where \ben
\bar{K} = - \na^2 + V''(\vph).
\een
Our Gaussian approximation can be unpacked further.  The
integration over $\eta$ may be re-expressed as
\ben
\int_{-\infty}^{+\infty} d\al D\eta\exp\(-\be\int_v \eta(\de
S_v/\de\vph+i\al) -
\half\be\int_v\eta\bar{K}\eta\),
\label{eEtaInt}
\een
which is proportional to
\ben
M_v(\vph) = \int_{-\infty}^{+\infty}d\al\exp\(\half\be\int_v(\de
S_v/\de\vph+i\al)\bar{K}^{-1} (\de S_v/\de\vph+i\al)\).
\een
The $\al$ integration fixes $M_v(\vph)$ through the expression
\ben
{1\over M_v^2(\vph)} = {1\over v}\int {d\bk \over
(2\pi)^3}{|I(\bk)|^2\over
\bk^2 + V''(\vph)}.
\label{eMsqInv}
\een
This is the high-temperature interacting field counterpart to
(\ref{hotint}).
Inserting (\ref{eMsqInv}) into (\ref{eEtaInt}) gives the simpler
expression
\ben
\exp\(\be v \om_v(j)\) \simeq N\int d\vph M_v(\vph) \exp\(\be
v[j\vph -
V(\vph)]\), \label{eEtoOm}
\een
renormalised to $\om_v(0)=0$. (The factor $M_v(\vph)$ was missed in
\cite{or}.)  $V(\vph)$ in (\ref{eEtoOm}) now contains the Gaussian
fluctuations of the light mode $\eta$.  However, as we observed
earlier, this will only affect non-leading behaviour in the high-$T$
expansion of $V(\vph)$, and we leave it unchanged.  Only when the
inequality (\ref{Gin}) is
satisfied will the contribution be noticeable, at which stage our
approximation breaks down, anyway.

An analytic estimate of $\om_v(j)$ requires further approximation.
We assume
that the $\vph$ integral in (\ref{eEtoOm}) is dominated by its
saddle-point at
$\vphh$ (remember $j>0$), which is
\ben
j=V'(\vphh)=\mu^2(T)\vphh+\la\vphh^3/3!,
\een
even when $v$ is not very large.  What makes this plausible is the
factor
$M_v(\vph)$.  Assume that we can approximate $M_v(\vph)$ by its
infinite volume
limit $\sqrt{V''(\vph)}$ near the saddle-point.  This then cancels
the first
non-leading order term in an expansion in powers of $(\be v)^{-1}$.
 Higher
terms are down by powers of $\la$.  The end result is that
\ben
\exp\(\be v \om_v(j)\) \simeq N\int d\vph
\surd(\mu^2(T)+\la\vph^2/2)\exp\(\be
v[j\vph+\mu^2(T)\vph^2/2+\la\vph^4/4!]\),
\een
for which
\ben
\om_v(j) \simeq j\vphh-\frac{1}{2}\mu^2(T)\vphh^2 -
\frac{1}{4!}\la\vphh^4.
\een
In turn,
\ben
V_v(\vph) \simeq \frac{1}{2}\mu^2(T)\vph^2 + \frac{1}{4!}\la\vph^4
\een
the required large volume result.  As we saw earlier, this
expression even gives the correct
qualitative behaviour for $v$=O$(|\mu(T)|^{-3})$, provided $\vph$ is
small enough, with
\ben
p(\vphv\ge\vph) \lap \exp\(-\vph^2/2T|\mu(T)|\),
\label{eAppProb}
\een
in agreement with (\ref{otm}).

We now turn to the problem at hand, the evaluation of the
probability bound in
the symmetry breaking phase, for which $V(\vph)$ is shown in Figure
1.
If $V''(\vph)$ is negative, the Gaussian
approximation that led to (\ref{eGauApp}) is unjustified and the
evaluation of
the $\eta$ integral is no longer possible in quite the same way.
However, it
is useful to decompose $\exp\(\be v \om_v(j)\)$ as
\ben
\exp\(\be v \om_v(j)\) \simeq \int_{C} d\vph M_v(\vph)\exp\(\be
v[j\vph-
V(\vph)]\) + {\rm the\ rest},
\een
where $C$ is the set of $\vph$ for which $V''$ is positive.

The relevance of this separation is that the saddle-points of the
exponent in
the first term occur when $V'' >0$.  Since the integrand
approximately vanishes at the points of inflexion, isolating
independent saddlepoints, it is tempting to approximate
further by
\ben
\exp\(\be v \om_v(j)\) \simeq \int_{C} d\vph\sqrt{V''(\vph)}
\exp\(\be v(j\vph-V(\vph))\)
\label{eRaysApp}
\een
where the saddle-points are the solutions to
\ben
0<j=V'(\vph) = \mu^2(T)\vph+\la\vph^3/3!.
\een
This relationship is displayed in Figure 2.  The factor of
$\sqrt{V''}$, which compensates for the Gaussian fluctuations,
guarantees
that all the saddle-points have equal weights.  Further, even when
$\be v$ is
not very large, the saddle-point contributions alone are accurate to
order
$(\be v)^{-2}$.

For small $j>0$ the saddle-point solutions occur at $\vph_{\pm} =
\pm\si+j/m^2(T)$, where $m^2 = -2\mu^2(T)$ is the scalar mass
squared in the
broken phase, and
$\si^2 = -2\mu^2(T)/\la$.  (Here, small means $j \ll m^2\si$.)  The
vanishing of the integrand at the points of inflection of $V$
guarantees that the saddle points are well separated, whence
equation  (\ref{eRaysApp}) gives
\ben
\exp\(\be v \om_v(j)\) \simeq \frac{1}{2}\exp\(\be v j^2/2m^2\)
\cosh(\be v \si j),
\label{eParFun}
\een
so that
\ben
\om_v(j) \simeq {j^2\over 2m^2} + {1\over \be v} \ln\(\cosh(\be
v \si
j)\).
\label{ovj}
\een
Our approximation to the integral (\ref{eRaysApp}) is good as long
as
$\be v\si^2m^2 \gg 1$, about which more will be said below.
It follows that
\ben
\vph = {d\om_v\over d j} \simeq {j\over m^2} + \si\tanh\be v\si
j,
\een
(the solid line in Figure 2).  As $v\to
\infty$, $\vph\to
\si$ (recall that $j/m^2 \ll \si$)
and the region $0<\vph<\si$ (the solid base of Figure 1) becomes
inaccessible.
 A small
$j$ expansion of $\om_v$ (small now also meaning $\be v\si j \ll 1$,
or equivalently $\vph^2 \ll \si^2$) gives
\ben
\om_v(j) \simeq \half j^2(\be v \si^2 +  m^{-2})
\een
with (convex) Legendre transform
\ben
V_v(\vph) \simeq \half\vph^2/(\be v\si^2+ m^{-2}).
\een
As $v\to\infty$ $V_v(\vph)$ becomes flat-bottomed (the dot-dashed
line in
the upper half of Figure 1, up to a constant).  However, the bound
on $p$
is not constant, but becomes
\ben
p(\vphv\ge\vph) \lap \exp\[-\half\vph^2/(\si^2(T)+1/m^2\be v)\],
\label{bigvbs}
\een
For large $v$ this is Gaussian and, approximately, {\it
independent\/}
of $v$.

For $\be v\si j \gg1$, or $\vph^2 \gg \si^2$,   there is only
one saddle-point, and insofar as this saddle-point
dominates, we
return to the form
\ben
p(\vphv \ge \vph) \lap \exp\(-\be vV(\vph)\).
\een

This is to be understood in the following way.  In the symmetric
phase we saw that the exponent of the $p$-bound (\ref{vbound}) was
approximately proportional
to $v$.  That is, the probability that the field average exceeds a
specified
value in a volume $v$ is equal to the probability that the field
exceeds this value in a unit volume to the power $v$.  Equivalently,
the field average behaves independently from one unit volume to the
next.
This dependence on the volume changes at $v \sim \mu^{-3}$,
consistent with the correlation length being the scale of the
fluctuations.  On this scale, the mean square fluctuations have
amplitude $T\mu$.  In the broken phase, we find that fluctuations
with $\vph  > \si $ are strongly damped beyond $(\vph-\si)^2 >
(\be v m^2)^{-1}$, and similarly for $\vph < -\si$.  For $\vph^2 \ll
\si^2$, there is a volume independent term in the probability.  This
has nothing to do with thermal fluctuations: it arises instead from
the correlations set up
by the field being in the broken phase where its expectation value
is
non-zero.  We see
from (\ref{eParFun}) that the partition function is the sum of two
pieces,
each with $\langle\vphb\rangle
= \pm\si$, and the applied current
weights them differently in order to obtain expectation values
of the field between the minima.   The probability bound is in fact
still determined by the mean square
fluctuations on a scale $v$, for as $j\to 0$,
\ben
\langle\vphv^2\rangle = {1\over \be v}{d^2\om_v\over dj^2} +\vph^2
\simeq
{1\over\be v m^2} + \si^2.
\een
We see clearly in this expression the separate effects of the
fluctuations and the correlations.

Our approximation to (\ref{eRaysApp}) goes wrong at $\be vm^2\si^2
\lap 1$,
or equivalently
\ben
1- {T^2\over T_c^2}\lap \la\({1\over m^3 v}\)^2.
\een
The most conservative bound on how near the critical temperature we
can go is obtained by taking $v$ to be a correlation volume,
and we recover
the Ginzburg criterion (\ref{Gin}).  Although taking $v\sim m^{-3}$
is
beyond the strict domain of applicability of our approximations, our
experience with the symmetric phase leads
us to hope that the true behaviour is  reproduced reasonably well.

\section{Discussion}

In this preparatory analysis of fluctuations a simple picture has
emerged.  Consider an arbitrary volume $v$ in which the field has an
average value $\vphi_v$.  Then the probability $p$, that $\vphi_v$
is not less than some chosen $\vphi$, is bounded above as
\begin{equation}
p(\vphv\ge\vphi) \le \exp(-\be vV_v(\vphi)).
\label{pbd}
\een
where $V_v(\phi)$ is a coarse-grained {\it convex} effective
potential.

In the {\it symmetric} phase $V_v(\vphi)$ is built from
coarse-grained Green functions in an obvious way.  To a good
approximation the bound is Gaussian, with independent fluctuations
in independent volumes.

In the {\it symmetry-broken} phase, when the one-loop effective
potential $V(\vphi)$ is non-convex, $V_v(\vphi)$ tends to the convex
hull $V_\infty(\vphi)$ of $V(\vphi)$ as $v\longrightarrow\infty$.
Although $V_\infty(\vphi)$ is flat-bottomed (the solid line in
Figure 1) the bound is still
approximately Gaussian, but now essentially independent of $v$.
This is a result of the thermal ensemble being comprised of
a sum of broken symmetry states with $\langle\vph\rangle =
\pm\si$.
In this context, the Ginzburg criterion (\ref{Gin}) is seen as the
condition that
ensuring that the probability bound is  vanishingly small for
fluctuations of magnitude $\sigma$ (see Figure 1) in
correlation-sized volumes.

Second-order transitions lack the dynamics of first-order
transitions; no tunnelling from metastable states, no bubble
nucleation.
The picture that we have presented is essentially static, somewhat
akin (in the symmetric phase, at least) to counting the area of
mountain-tops that protrude above a given cloud level as a fraction
of the total area.  Of course, fluctuations come and go, and what we
are seeing is the statistical average.  However, this same
statistical sampling will be very useful for first-order
transitions, enabling us to estimate the fraction of space populated
by fields near the global minimum, prior to accounting for
tunnelling.  This will be discussed elsewhere.

We wish to thank Urs Wiedeman for making reference \cite{jona} known
to us.  MH is supported by the SERC.

\section*{Figure Captions}

Figure 1.  The thermal effective potential $V(\vphi)$ in the
symmetry-broken phase.  The solid line denotes the infinite volume
convex hull $V_{\infty}(\vphi)$ of $V(\vphi)$.

Figure 2.  The relationship between $\vphi$ and $j$ in the broken
phase.  The dashed line
identifies the saddlepoints.  The solid line
shows the interference between saddlepoints.

\vfill\eject

\vfill\eject
\section{Appendix}
In this appendix we evaluate the distribution flunctional
$\Ga[\eta,\vph_0]$,
whose exponential weights the fluctuations around a background field
$\vphb = \vph_0 - \eta$, so that
\ben
P[\vphb] = \int D\eta \exp\(-\be\Ga[\eta,\vph_0] - \be
\seff[\vph_0]\).
\een
We recall that $\Ga$ is defined as the Legendre transform of
$\Om[\{J_n\},\vph_0]$, the generator of connected heavy mode Green's
functions, evaluated at $J_n = i\La$.  That is,
\ben
\Ga[\eta,\vph_0] = \Om[i\La,\vph_0] - i\int \La\eta,
\een
with
\ben
\frac{\de\Om}{\de(i\La)} = \eta.
\een
Let us now evaluate $\Ga$ for our interacting scalar theory with
$V(\phi) = \frac{1}{2}\mu^2\phi^2 + \frac{1}{4!}\la\phi^4$.  First
we
define a heavy mode propagator $\De_n(\bx,\by) = \de(\bx-\by)
(-\na_x^2+m_n^2)^{-1}$, where
$m_n^2 = \mu_0^2 + \half\la\vph_0^2+ (2\pi n T)^2$.  Then
\bea
-\Om[i\La,\vph_0] &=& \int(i\La)K^{-1}(i\La) +
\frac{\la}{3!}{\sum_{n_1,n_2}}'\int\vph_0\De_{n_1}
\De_{n_2}\De_{-n_1-n_2}(i\La)^3 \\
&&+ \frac{\la}{4!}{\sum_{n_1,n_2,n_3}}^{\kern-5.5pt\prime}\int
\De_{n_1}\De_{n_2}\De_{n_3}\De_{-n_1-n_2-n_3},
(i\La)^4,
\eea
where $\sum'_{n_1,n_2}$ denotes sums excluding zero and also
excluding
$n_1+n_2=0$ (and similarly for three arguments).
The fluctuation functional has the expansion
\ben
\Ga[\eta,\vph_0] = \frac{1}{2} \int\Ga^{(2)}\eta^2 +
\frac{1}{3!}\int\Ga^{(3)}\eta^3 + \frac{1}{4!}\int\Ga^{(4)}\eta^4 +
\cdots,
\een
where $\Ga^{(2)} = K = \sum_n\De_n$.  In the high T limit the
expressions
simplify considerably, for $K^{-1} \to \be^2/12$.  Then we find
\bea
\Ga^{(3)} &\simeq& G^{(3)}(\be^2/2)^{-3} \simeq
\la\vph_0\({6\over\pi^2}\)^3
{\sum_{n_1,n_2}}'{1\over n_1^2n_2^2(n_1+n_2)^2},\\
\Ga^{(4)} &\simeq& G^{(4)}(\be^2/12)^{-4} \simeq \la
\({6\over\pi^2}\)^4
{\sum_{n_1,n_2,n_3}}^{\kern-5.5pt\prime}
{1\over n_1^2n_2^2n^2_3(n_1+n_2+n_3)^2}.
\eea
We do not evaluate the sums,as they are inessential for our
purposes.
We merely denote them $\si_a$ and $\si_b$ respectively, to arrive at
\ben
\Ga[\eta,\vph_0] = \frac{1}{2}\int K\eta^2 +
\frac{\la}{3!}\int\vph_0\eta^3\si_a + \frac{\la}{4!}\int\eta^4\si_b.
\een

\end{document}